



\font\twelverm=cmr10 scaled 1200    \font\twelvei=cmmi10 scaled 1200
\font\twelvesy=cmsy10 scaled 1200   \font\twelveex=cmex10 scaled 1200
\font\twelvebf=cmbx10 scaled 1200   \font\twelvesl=cmsl10 scaled 1200
\font\twelvett=cmtt10 scaled 1200   \font\twelveit=cmti10 scaled 1200

\skewchar\twelvei='177   \skewchar\twelvesy='60


\def\twelvepoint{\normalbaselineskip=12.4pt
  \abovedisplayskip 12.4pt plus 3pt minus 9pt
  \belowdisplayskip 12.4pt plus 3pt minus 9pt
  \abovedisplayshortskip 0pt plus 3pt
  \belowdisplayshortskip 7.2pt plus 3pt minus 4pt
  \smallskipamount=3.6pt plus1.2pt minus1.2pt
  \medskipamount=7.2pt plus2.4pt minus2.4pt
  \bigskipamount=14.4pt plus4.8pt minus4.8pt
  \def\rm{\fam0\twelverm}          \def\it{\fam\itfam\twelveit}%
  \def\sl{\fam\slfam\twelvesl}     \def\bf{\fam\bffam\twelvebf}%
  \def\mit{\fam 1}                 \def\cal{\fam 2}%
  \def\tt{\twelvett}
  \textfont0=\twelverm   \scriptfont0=\tenrm   \scriptscriptfont0=\sevenrm
  \textfont1=\twelvei    \scriptfont1=\teni    \scriptscriptfont1=\seveni
  \textfont2=\twelvesy   \scriptfont2=\tensy   \scriptscriptfont2=\sevensy
  \textfont3=\twelveex   \scriptfont3=\twelveex  \scriptscriptfont3=\twelveex
  \textfont\itfam=\twelveit
  \textfont\slfam=\twelvesl
  \textfont\bffam=\twelvebf \scriptfont\bffam=\tenbf
  \scriptscriptfont\bffam=\sevenbf
  \normalbaselines\rm}



\def\beginlinemode{\endmode
  \begingroup\parskip=0pt \obeylines\def\\{\par}\def\endmode{\par\endgroup}}
\def\beginparmode{\endmode
  \begingroup \def\endmode{\par\endgroup}}
\let\endmode=\par
{\obeylines\gdef\
{}}
\def\singlespace{\baselineskip=\normalbaselineskip}

\def\oneandahalfspace{\baselineskip=\normalbaselineskip
  \multiply\baselineskip by 3 \divide\baselineskip by 2}
\def\doublespace{\baselineskip=\normalbaselineskip \multiply\baselineskip by 2}

\newcount\firstpageno
\firstpageno=2
\footline={\ifnum\pageno<\firstpageno{\hfil}\else{\hfil\twelverm\folio\hfil}\fi}
\let\rawfootnote=\footnote		
\def\footnote#1#2{{\rm\singlespace\parindent=0pt\rawfootnote{#1}{#2}}}
\def\raggedcenter{\leftskip=4em plus 12em \rightskip=\leftskip
  \parindent=0pt \parfillskip=0pt \spaceskip=.3333em \xspaceskip=.5em
  \pretolerance=9999 \tolerance=9999
  \hyphenpenalty=9999 \exhyphenpenalty=9999 }
\def\dateline{\rightline{\ifcase\month\or
  January\or February\or March\or April\or May\or June\or
  July\or August\or September\or October\or November\or December\fi
  \space\number\year}}
\def\received{\vskip 3pt plus 0.2fill
 \centerline{\sl (Received\space\ifcase\month\or
  January\or February\or March\or April\or May\or June\or
  July\or August\or September\or October\or November\or December\fi
  \qquad, \number\year)}}


\hsize=6.5truein
\hoffset=0truein
\vsize=8.9truein
\voffset=0truein
\parskip=\medskipamount
\twelvepoint		
\doublespace		
\overfullrule=0pt	


\def\preprintno#1{
 \rightline{\rm #1}}	

\def\title			
  {\null\vskip 3pt plus 0.2fill
   \beginlinemode \doublespace \raggedcenter \bf}

\def\author			
  {\vskip 3pt plus 0.2fill \beginlinemode
   \singlespace \raggedcenter}

\def\affil			
  {\vskip 3pt plus 0.1fill \beginlinemode
   \oneandahalfspace \raggedcenter \sl}

\def\abstract			
  {\vskip 3pt plus 0.3fill \beginparmode
   \doublespace \narrower ABSTRACT: }

\def\endtitlepage		
  {\endpage			
   \body}

\def\body			
  {\beginparmode}		

\def\head#1{			
  \filbreak\vskip 0.5truein	
  {\immediate\write16{#1}
   \raggedcenter \uppercase{#1}\par}
   \nobreak\vskip 0.25truein\nobreak}

\def\subhead#1{			
  \vskip 0.25truein		
  {\raggedcenter #1 \par}
   \nobreak\vskip 0.25truein\nobreak}

\def\refto#1{$^{#1}$}		

\def\references			
  {\head{References}		
   \beginparmode
   \frenchspacing \parindent=0pt \leftskip=1truecm
   \parskip=8pt plus 3pt \everypar{\hangindent=\parindent}}

\gdef\refis#1{\indent\hbox to 0pt{\hss#1.~}}	

\gdef\journal#1, #2, #3, 1#4#5#6{		
    {\sl #1~}{\bf #2}, #3 (1#4#5#6)}		

\gdef\journ2 #1, #2, #3, 1#4#5#6{		
    {\sl #1~}{\bf #2}: #3 (1#4#5#6)}		

\def\refstylenp{		
  \gdef\refto##1{ [##1]}				
  \gdef\refis##1{\indent\hbox to 0pt{\hss##1)~}}	
  \gdef\journal##1, ##2, ##3, ##4 {			
     {\sl ##1~}{\bf ##2~}(##3) ##4 }}

\def\refstyleprnp{		
  \gdef\refto##1{ [##1]}				
  \gdef\refis##1{\indent\hbox to 0pt{\hss##1)~}}	
  \gdef\journal##1, ##2, ##3, 1##4##5##6{		
    {\sl ##1~}{\bf ##2~}(1##4##5##6) ##3}}

\def\prd{\journal Phys. Rev. D, }

\def\prl{\journal Phys. Rev. Lett., }

\def\pl{\journal Phys. Lett., }

\def\endreferences{\body}

\def\figurecaptions		
  {\endpage
   \beginparmode
   \head{Figure Captions}
}

\def\endfigurecaptions{\body}

\def\endpage			
  {\vfill\eject}

\def\endpaper			
  {\endmode\vfill\supereject}

\def\endit
  {\endpaper\end}


\def\ref#1{Ref. #1}			
\def\Ref#1{Ref. #1}			

\def\frac#1#2{{\textstyle #1 \over \textstyle #2}}

\def\sla{\raise.15ex\hbox{$/$}\kern-.57em}
\def\leaderfill{\leaders\hbox to 1em{\hss.\hss}\hfill}
\def\twiddle{\lower.9ex\rlap{$\kern-.1em\scriptstyle\sim$}}
\def\bigtwiddle{\lower1.ex\rlap{$\sim$}}
\def\gtwid{\mathrel{\raise.3ex\hbox{$>$\kern-.75em\lower1ex\hbox{$\sim$}}}}
\def\ltwid{\mathrel{\raise.3ex\hbox{$<$\kern-.75em\lower1ex\hbox{$\sim$}}}}
\def\square{\kern1pt\vbox{\hrule height 1.2pt\hbox{\vrule width 1.2pt\hskip 3pt
   \vbox{\vskip 6pt}\hskip 3pt\vrule width 0.6pt}\hrule height 0.6pt}\kern1pt}

\catcode`@=11
\newcount\r@fcount \r@fcount=0
\newcount\r@fcurr
\immediate\newwrite\reffile
\newif\ifr@ffile\r@ffilefalse
\def\w@rnwrite#1{\ifr@ffile\immediate\write\reffile{#1}\fi\message{#1}}

\def\writer@f#1>>{}
\def\referencefile{
  \r@ffiletrue\immediate\openout\reffile=\jobname.ref%
  \def\writer@f##1>>{\ifr@ffile\immediate\write\reffile%
    {\noexpand\refis{##1} = \csname r@fnum##1\endcsname = %
     \expandafter\expandafter\expandafter\strip@t\expandafter%
     \meaning\csname r@ftext\csname r@fnum##1\endcsname\endcsname}\fi}%
  \def\strip@t##1>>{}}

\def\citeall#1{\xdef#1##1{#1{\noexpand\cite{##1}}}}
\def\cite#1{\each@rg\citer@nge{#1}}	

\def\each@rg#1#2{{\let\thecsname=#1\expandafter\first@rg#2,\end,}}
\def\first@rg#1,{\thecsname{#1}\apply@rg}	
\def\apply@rg#1,{\ifx\end#1\let\next=\relax
\else,\thecsname{#1}\let\next=\apply@rg\fi\next}

\def\citer@nge#1{\citedor@nge#1-\end-}	
\def\citer@ngeat#1\end-{#1}
\def\citedor@nge#1-#2-{\ifx\end#2\r@featspace#1 
  \else\citel@@p{#1}{#2}\citer@ngeat\fi}	
\def\citel@@p#1#2{\ifnum#1>#2{\errmessage{Reference range #1-#2\space is bad.}%
    \errhelp{If you cite a series of references by the notation M-N, then M and
    N must be integers, and N must be greater than or equal to M.}}\else%
 {\count0=#1\count1=#2\advance\count1
by1\relax\expandafter\r@fcite\the\count0,%
  \loop\advance\count0 by1\relax
    \ifnum\count0<\count1,\expandafter\r@fcite\the\count0,%
  \repeat}\fi}

\def\r@featspace#1#2 {\r@fcite#1#2,}	
\def\r@fcite#1,{\ifuncit@d{#1}
    \newr@f{#1}%
    \expandafter\gdef\csname r@ftext\number\r@fcount\endcsname%
                     {\message{Reference #1 to be supplied.}%
                      \writer@f#1>>#1 to be supplied.\par}%
 \fi%
 \csname r@fnum#1\endcsname}
\def\ifuncit@d#1{\expandafter\ifx\csname r@fnum#1\endcsname\relax}%
\def\newr@f#1{\global\advance\r@fcount by1%
    \expandafter\xdef\csname r@fnum#1\endcsname{\number\r@fcount}}

\let\r@fis=\refis			
\def\refis#1#2#3\par{\ifuncit@d{#1}
   \newr@f{#1}%
   \w@rnwrite{Reference #1=\number\r@fcount\space is not cited up to now.}\fi%
  \expandafter\gdef\csname r@ftext\csname r@fnum#1\endcsname\endcsname%
  {\writer@f#1>>#2#3\par}}

\def\ignoreuncited{
   \def\refis##1##2##3\par{\ifuncit@d{##1}%
     \else\expandafter\gdef\csname r@ftext\csname
r@fnum##1\endcsname\endcsname%
     {\writer@f##1>>##2##3\par}\fi}}

\def\r@ferr{\endreferences\errmessage{I was expecting to see
\noexpand\endreferences before now;  I have inserted it here.}}
\let\r@ferences=\references
\def\references{\r@ferences\def\endmode{\r@ferr\par\endgroup}}

\let\endr@ferences=\endreferences
\def\endreferences{\r@fcurr=0
  {\loop\ifnum\r@fcurr<\r@fcount
    \advance\r@fcurr by 1\relax\expandafter\r@fis\expandafter{\number\r@fcurr}%
    \csname r@ftext\number\r@fcurr\endcsname%
  \repeat}\gdef\r@ferr{}\endr@ferences}


\let\r@fend=\endpaper\gdef\endpaper{\ifr@ffile
\immediate\write16{Cross References written on []\jobname.REF.}\fi\r@fend}

\catcode`@=12

\citeall\refto		
\citeall\ref		%
\citeall\Ref		%

\def\caption#1{\centerline{
	\uppercase{#1}}\noindent}
\newdimen\psfigsize
\newcount\psfigcount
\def\psfigure#1 #2 #3 #4 #5{\topinsert\vbox{
    \psfigcount=#1
    \psfigsize=#1truept
    \vskip \psfigsize
    \includegraphics{#4}
    \vskip 10truept
    \caption {#3}
    {\narrower\singlespace\noindent#5\par}\vskip 0.1truein
    plus0.2truein}
\endinsert}
%
%
%
\def\psoddfigure#1 #2 #3 #4 #5 #6{\topinsert\vbox{
    \psfigcount=#1
    \psfigsize=#2truept
    \vskip \psfigcount truept
    \includegraphics{#5}
    \advance\psfigsize by -\psfigcount truept \vskip\psfigsize
    \vskip 10truept
    \caption {#4}
    {\narrower\singlespace\noindent#6\par}\vskip 0.1truein
    plus0.2truein}
\endinsert}
%
\def\figurespace#1 #2 #3 #4 {\topinsert\vbox{
    \psfigcount=#1
    \psfigsize=#1truept
    \vskip \psfigsize
    \vskip 10truept
    \caption {#3}
    {\narrower\singlespace\noindent#4\par}\vskip 0.1truein
    plus0.2truein}
\endinsert}

\def\preprintno#1{
 \rightline{\rm #1}}	

\def\TABLEcap#1#2{
 {\narrower\medskip\singlespace\noindent{ Table~#1.}#2 \smallskip
   \par
}}

\def\dbline{\noalign{\hrule}\noalign{\vskip 2pt}\noalign{\hrule}}
\def\sgline{\noalign{\hrule}}
\def\notext{\omit&\omit&\omit&\omit&\omit&\omit\cr}

\def\fstrut{\vrule height 13.5pt depth 2.25pt width 0pt}

\def\endrule{&\omit\fstrut\vrule\cr}


\def\preprint{N}        

\def\Dslash{\rlap{\bf D}\kern1.5pt/\kern1.5pt}
\vskip -.5truein
{\baselineskip=12pt
\preprintno{ COLO-HEP-282}
}
\vbox to 3.1truein {
\title
MORE ABOUT   ORBITALLY EXCITED HADRONS FROM   LATTICE QCD
\author
Thomas A. DeGrand
\author
Matthew W. Hecht
\affil
Department of Physics
University of Colorado
Boulder, Colorado 80309
\endgroup
\vfill
}
\abstract
This is a second paper describing the calculation of spectroscopy for
orbitally excited states from lattice simulations of Quantum Chromodynamics.
New features include higher statistics for P-wave systems
and first results for the spectroscopy of D-wave mesons
and baryons,  for relatively heavy quark masses.
We parameterize the Coulomb gauge wave functions for P-wave and D-wave
systems and compare them  to those of their corresponding S-wave states.
\endpage
\body
\head{I. Introduction}
Recently we presented some preliminary results from
 an exploratory calculation of the masses
of P-wave mesons and baryons from lattice Monte Carlo simulations
of Quantum Chromodynamics in quenched approximation with Wilson
fermions\rlap.\refto{TDMHPWAVE}
This communication describes the final results from these simulations.
 We have increased the statistics of our P-wave
study. There is a hint of some fine structure splitting in the
charmonium system. We also present first results for D-wave spectroscopy
of fairly heavy quark mesons and baryons. Finally, we show some of the
properties of Coulomb gauge meson and baryon wave functions of
P-wave and D-wave systems, and compare them to S-wave wave functions
at the same quark masses\rlap.\refto{TDMHSWAVE}

Only recently has lattice QCD spectroscopy begun to move beyond
 ground state hadrons. Some P-wave states' masses are regularly measured in
staggered simulations because they are the odd parity partners of
``ordinary'' states: the $a_1$ and $\rho$ are examples of such pairs.
In  nonrelativistic
QCD, Lepage and Thacker\refto{LEPAGET} have computed the masses of $\chi_C$
and $\chi_B$ states (without including spin effects).
Few Wilson simulations have studied P-wave states.
The APE collaboration\refto{APE}  measured masses of
 some P-wave mesons in quenched simulations at $6/g^2 =\beta=5.7$, but
has had difficulty in continuing their program to higher
 $\beta$\rlap.\refto{APETWO} Recently, El-Khadra, Hockney, Kronfeld, and
Mackenzie have presented a calculation of the 1P-1S splitting in
charmonium, which they use to fix the strong coupling
constant\rlap.\refto{FNAL}
This calculation was done with a smaller lattice spacing
than the one we report here, and with an improved action for the fermions.

Calculations of the masses of orbital excitations
in lattice simulations are difficult for three reasons: First, one needs
to measure a correlation function
 with nonzero overlap onto the desired $L$ sector and zero
overlap on $L=0$, otherwise the signal will be dominated at large $t$ by
the lighter $L=0$ states. Our methodology solves this problem.
  Second, the signal is intrinsically noisy\rlap.\refto{LEPAGE}
A diagonal correlator,
$C(t) = \langle 0 | \Gamma(t) \Gamma(0) | 0\rangle ,$
which falls off at large $t$ like
$\exp(-E_1t)$
where $E_1$ is the energy of the lightest state which the operator $\Gamma$
can create from the vacuum, has
fluctuations  given by
$$\sigma^2_\Gamma = {1 \over N} (\langle |\Gamma(t) \Gamma(0)|^2\rangle
 - C(t)^2 ). \eqno(1.1)$$
Due to its first term,  $\sigma_\Gamma^2$
 decays with a mass characteristic of the lightest particle $|\Gamma|^2$ can
make from the vacuum.
 If $\Gamma$ is a meson operator (creating a $\bar q q$ pair)
$\Gamma^2$ will create a $q^2 \bar q ^2$ state,  which will
most likely  couple to a $\pi\pi$ pair.  Its correlator
will fall like $\exp(-2m_\pi t)$.  In the baryon sector $|\Gamma|^2$ will make
 a $q^3 \bar q^3$ state, and the lightest such state is three pions.
Thus we expect a signal to noise ratio to be a falling function of $t$:
$\sigma/C_H(t) \simeq \exp{(m_H - m_\pi)}t$
for mesons, and
$\sigma/C_H(t) \simeq \exp{(m_H - 3/2m_\pi)}t $
for baryons.
This is a  more serious problem for
orbitally
excited states than for S-wave states  because their energy differences
 are larger.
 Finally,  the baryon sector includes multiple
 states with the same quantum numbers, which will appear in
the same correlators.  For example,
in the $L=1$ [70] of SU(6)\rlap,\refto{KOKKEDDEE} the
nonstrange  sector includes  one
$j=5/2$ and two  $j=3/2$ and $j=1/2$ nucleons, and $j=3/2$ and $j=1/2$
$\Delta$'s.

\head{II. Methodology}

\subhead{A. Construction of Orbitally Excited States}

We construct orbitally excited states by using
interpolating fields which couple only to a specific angular momentum
eigenstate, which are projected onto zero momentum and which
 are of large spatial extent to maximize overlap with the state.
Our strategy is look at correlators of different operators at $t=0$
and $t \neq 0$.

At the $t \neq 0$ end
of the correlation function
we use an operator which depends on the relative separation of the
quarks, which is conventionally referred to as
a ``wave function\rlap.\refto{ALLWF}''
The  wave function $\psi_G(r)$ of a hadron H in a gauge G
 is defined as
$$\psi_G(r) =
\sum_{\vec x} \langle H | q(\vec x)
{\bar q}(\vec x + \vec r) | 0 \rangle \eqno(2.1)
$$
where $q(\vec x)$ and
${\bar q}(\vec y)$ are quantum mechanical operators
which create a quark and an antiquark at locations
$\vec x$ and $\vec y$. (We have suppressed Dirac and color indices.)
Our correlation function is  constructed from
 convolutions of quark and antiquark propagators
$G(x,y)$
$$C(\vec r,t) = \sum_{\vec x}\langle 0 |\Psi(\vec y_1, \vec y_2)
G_q(\vec y_1,0;\vec x,t)
G_{\bar q}(\vec y_2,0;\vec x + \vec r,t)
|0 \rangle \eqno(2.2)$$
where $\Psi(\vec y_1, \vec y_2)$ is the $t=0$ operator.
At large $t$ if the mass of the hadron is $m_H$, then
$$C(\vec r,t ) \simeq \exp (-m_H t) \psi_G(\vec r) \eqno(2.3)$$
and so by plotting $C(\vec r, t)$ as a function of $\vec r$ we can reconstruct
the wave function up to an overall constant.
We measure the mass of a state by convoluting $C(\vec r,t)$ with some
  test function which further projects out the desired
state:
$$C(t) = \sum_{\vec r} \psi^*_{test}(\vec r) C(\vec r,t) .\eqno(2.4)$$

At $t=0$ we take an operator which is separable in the coordinates
of the quarks. For a meson we use
$$\Psi(x_1,x_2) = \phi_1(x_1) \phi_2(x_2). \eqno(2.5) $$
In order to couple to orbital excitations we take $\phi_1$
to be  an S-wave and $\phi_2$ to be some orbitally excited state
with angular momentum $l$,
 centered around some
specified coordinate.
 This state is a
linear superposition of a $\vec p = 0$ $L=l$ orbital excitation and a state
whose center of mass momentum is nonzero
 (this is the familiar ``translation mode'' of a shell model state).
Convoluting quark propagators as
in Eqn. (2.2) removes the $\vec p \neq 0$ state and gives us
the wave function of the $\vec p = 0$ $L=l$ state.

\subhead{B. Spin Considerations}

We did not construct
a complete set of P-wave  or D-wave mesons and baryons.
Instead, we proceeded as follows (for P-waves):
We worked in a basis in which $\gamma_0$ is diagonal. Our sources and
sinks were chosen to couple only to the upper (large) components
of the Dirac spinor. We constructed propagators for S-wave quarks with
$m_s= \pm 1/2$ and for P-wave quarks with $m_l=1$, $m_s= \pm 1/2$.
We can then completely construct the $|jm \rangle = |22\rangle$ ${}^3P_2$
and $|jm \rangle = |11\rangle$ ${}^1P_1$ mesons, as well as the
$|jm \rangle = |{5 \over 2} {5 \over 2} \rangle$ nucleon $N(5/2)$.
We formed two other  meson states with  $S=1$ and $m_j$ = 1 and 0
($m_S = 0$ and -1)
which are not angular momentum eigenstates;  while we will
label them as ${}^3P_1$ and ${}^3P_0$ they couple to  all three ${}^3P$
states.
(The ``${}^3P_1$'' state is
$(\uparrow\downarrow + \downarrow\uparrow) | 11 \rangle$; the
``${}^3P_0$'' state is
$(\downarrow\downarrow ) | 11 \rangle$.)
If the ordering of states on the lattice were as in charmonium, the lightest
state in the channel would have the smallest $j$ for a given $m_j$ and the
 wavefunctions would in fact project out the states which they label.
 In the baryon sector we constructed $uud$ bound states
with $m_j = $ 3/2 $(\uparrow \uparrow\downarrow)|11\rangle$
and -1/2 $(\downarrow \downarrow\downarrow)|11\rangle$.

D-wave spectroscopy is identical apart from the substitution of
$l=2$ for $l=1$: we completely construct the ${}^3D_3$ and ${}^1D_2$
meson states, and construct $m_j=2$ and 1 states which overlap
${}^3D_2$ and ${}^3D_1$ states. We construct nucleons of $m_j=7/2$,
3/2 and 1/2.

This is an incomplete construction forced on us by computer memory
limitations and a desire to keep the calculation simple.
A dedicated simulation should do this properly, but will need many more quark
propagators.
Note also that while we are using nonrelativistic wave functions, they
have the quantum numbers appropriate to the desired
 states and will couple to them,
and not to S-wave states, regardless of whether the quarks are actually
relativistic.

\subhead{C. Details of the Simulations}

We performed the simulations using the Connection Machine CM-2 at the
Pittsburgh Supercomputing Center.
Our data set consists of eighty lattices
for P-wave studies, of which the last fifty were also
analyzed for D-wave systems, computed in quenched approximation
at coupling $\beta=6$, separated  by  a combination of microcanonical
 overrelaxed\refto{RELAX} and Kennedy-Pendleton quasi-heat bath\refto{KPQ}
 sweeps
(100 passes of a pattern of four overrelaxed sweeps through
the lattice and one heat bath sweep).
The lattice size is $16^4$ sites.  We gauge fixed  each of
the lattice configurations to
Coulomb  gauge using an overrelaxation algorithm\rlap.\refto{OVERRELAX}
We used Wilson fermions.
 We used a fast
matrix inverter provided by C. Liu of Thinking Machines, Inc\rlap.\refto{LIU}
to construct quark propagators. We computed P-wave spectroscopy
at three hopping parameters corresponding to relatively heavy quark masses:
$\kappa = 0.130$, 0.1450, and 0.1520,
and D-wave spectroscopy at $\kappa=0.1300$ and 0.1450.
The pseudoscalar mass in lattice
units at these three
$\kappa$'s is 1.43, 0.83, and 0.56.
 We used heavy quarks because the size of the wave
function for the orbitally excited states are
 much larger than the size of an S-wave
 bound state and if the quark mass becomes too small, the wave function
is squeezed by the simulation volume.

 In Eqn. (2.4)
we use a test function $\psi_{test} = \exp(-(r/r_0)^2) r^l Y_l^l(\Omega)$,
in the
relative coordinate.
We  take $\phi_1$ in Eqn. (2.5) to be a Gaussian centered at the origin
 and
$\phi_2$ to be $\exp(-(r/r_0)^2) r^l Y_l^l(\Omega)$.
(As a technical note, when the  width of the Gaussian  becomes large compared
to the size of the lattice one must worry about edge effects. We do that
by replacing the spherical harmonic by a function which has the same
symmetry but is periodic in the box (of size $L$):
 $Y_1^1 = (x + iy)/r \rightarrow (\sin  2 \pi x/L + i \sin 2 \pi y/L)/r$.)
 The width of the Gaussians used in $\Psi$ and $\psi_{test}$ was taken
to be 2, 2, and 2$\sqrt 2$ lattice spacings for the three $\kappa$'s.

We recorded wavefunctions for ${}^3P_2$, ${}^1P_1$, ${}^3D_3$
and ${}^1D_2$ mesons and for $N(5/2)$ and $N(7/2)$ baryons, at time slices
4, 5, and 6.  All baryon wave functions pin the two quarks in a relative
S-state
to the same coordinate.  We folded meson data onto one octant of the
spatial lattice and baryon data onto one quadrant before storing it,
and in addition kept data on one plane without folding.

\head{III. Wave Functions}

\subhead{A. Pictures of Wave Functions}
Now we   display some of the features of P-wave and D-wave
wave functions and compare some of their simple observables to those of
S-wave mesons of our earlier study\rlap.\refto{TDMHSWAVE}
We begin our display of results for wave functions
 by showing some pictures of Coulomb gauge
  meson and baryon
wave functions.

In Figs. 1 and 2 we display plots of the real part of the wave
 function in the plane $z=0$ for a $\kappa=.152$ $N(5/2)$ baryon
and a $\kappa=.145$ ${}^3D_3$ meson, respectively. The contours show
the locations where $\psi$ is a multiple of twenty per cent of its maximum
value. The data for these graphs have not been spatially averaged.
They show the characteristic dipole and quadrupole structure
of the appropriate spherical harmonic.
The fact that these distributions are not symmetric gives
the reader an impression of the fluctuations in the data.
They also give some idea
of the extent to which the granularity of the lattice
distorts the state, and the extent to which the state fits into
the simulation volume.

\if\preprint Y \psoddfigure  324 276 60 {Figure 1} {n52_152.ps} {
Wave function of the $\kappa=0.152$
$N(5/2)$ baryon in the plane $z=0$. The contours
show interpolated lines of constant real $\psi$ in
 multiples of 0.2 times the maximum value of $\psi$.
The shading shows the value of the wave function (black is most negative,
white most positive), interpolating from the original $16^2$ lattice
to a $64^2$ grid.}
\fi

\if\preprint Y \psoddfigure  324 276 60 {Figure 2} {d33_145.ps} {
Wave function of the $\kappa=.145$ ${}^3D_3$ meson in the plane $z=0$. The
 contours and shading are parameterized as in Fig. 1.}
\fi

In order to further illustrate the extent to which a state fits
in the simulation volume, we display in Fig. 3 a set of
three dimensional contours of constant absolute value
of the real part of the wave function. This figure,
for the $\kappa=.145$ ${}^3D_3$ meson, shows the characteristic
four lobe quadrupole structure whose outer regions are compromised by the
simulation volume. This is a problem for all the lighter mass
states; all the $\kappa=.130$ mesons appear to fit reasonably well into
the simulation volume.

\if\preprint Y \psoddfigure 522 486 -6 {Figure 3} {n52_152.ps} {
Surfaces of constant absolute amplitude of the
$\kappa=.145$ ${}^3D_3$ meson in
three dimensions, in fractions of its maximum: (a) $|\psi|= 0.2\psi_{max}$,
 (b) $|\psi|= 0.4\psi_{max}$ , (c) $|\psi|= 0.5\psi_{max}$,
 (d) $|\psi|= 0.6\psi_{max}$. The ``breaks'' in the surfaces in (a)
and (b) occur when the surfaces intersect the edge of the
simulation volume.}
\fi

Finally, we extract the radial wave function  $f(r)$ itself, from
$$f(r) = Re \ \psi(\vec r)/ Re Y_l^l(\Omega).  \eqno(3.1)$$
The Coulomb  gauge radial wave functions for $\kappa=.145$
${}^3P_2$ and ${}^3D_3$ mesons are shown in Figs. 4 and 5.
 They appear to show the characteristic
linear and quadratic zeros
of the  wave function
 at the origin.
\if\preprint Y \psoddfigure 324 324 60 {Figure 4} {fr_3p2_145.ps} {
Radial wave function of the $\kappa=.145$  ${}^3P_2$ meson.
Crosses show all points; squares are points along the  $x$ axis
used to fit the wave function (fit is the line).}
\fi

\if\preprint Y \psoddfigure 324 324 60 {Figure 5} {fr_3d3_145.ps} {
Radial wave function of the $\kappa=.145$  ${}^3D_3$ meson.
Crosses show all points; squares are points
used to fit the wave function (fit is the line).}
\fi

\subhead{B. Fitting Wave Function Parameters}
The goal of this section is to provide simple analytical parameterizations
of wave functions which can be used for future studies of spectroscopy
and to provide checks for calculations of wave function properties
performed directly on the data.

Fitting the wave functions proved to be unexpectedly difficult because
of the high correlations among wave functions at different separations.
With only fifty or eighty lattices we had to fit a subset of the data
since correlated fits require more lattices than fitted points.
We elected to choose coordinates along axes where the spherical
harmonic was unity (up to a sign); this gives us seven ($z=1$ to 7)
points to fit. We folded all directions related by reflections onto this
axis with the necessary signs.  Then we fit the radial wave function
$f(r)$ including a periodic or antiperiodic image term from the
boundary (we fit $f_{test}(r) = g(r) \pm g(L-r)$).

After a certain amount of trial and error we chose to fit to
$$g(r) = A (r + b r^2) \exp(-cr)  \eqno(3.2)$$
for P-waves and
$$g(r) = A (r^2 + b r^4) \exp(-cr)  \eqno(3.3)$$
for D-waves,
plus image terms,
and we believe that many other simple functional forms would work as well.
The data cannot distinguish between these or more complicated functions,
and if we try to force a fit, the Hessian matrix becomes singular.
We fit the baryon data to the same parameters (recall that we have
pinned two quarks together, so there is only one relative coordinate left).

The data are very correlated and the matrix of correlations very singular.
It was not unusual to find correlation matrices whose conditioning number
(ratio of largest to smallest eigenvalue) approached several thousand.
The conditioning number was not stable: fitting half the lattices in a data set
could cause the conditioning number to vary by a factor of two.
We could also achieve a considerable variation in the conditioning
number by varying elements in the correlation matrix by hand by a per cent
or so.
In contrast, the correlation matrices for propagators had conditioning numbers
on the order of 50 and were quite stable under the same tests.
 Typically, in a 7 by 7 correlation matrix, only
the largest 3 or 4 eigenvalues remained reasonably stable as the number
of lattices in the data set was varied. Therefore we adopted the following
strategy for determining the parameters in $f(r)$: we looked at
uncorrelated fits,  correlated fits to all parameters (very unstable),
correlated fits in which the correlation matrix had its smallest
three eigenvalues removed (via singular value decomposition)
and correlated fits to a subset of the data (often $r=1$,3,5,7).
In the latter case one could not use consecutive points since either
one or both of
the correlation matrix  or the Hessian matrix would
become singular. The D-wave data was much more difficult to deal with
than the P-wave data in this respect.

The overall normalization of the wave function is not important
for spectroscopy studies. The parameters $b$ and $c$ for
P-wave mesons are displayed in Fig. 6 and for D-wave mesons
in Fig. 7.
\if\preprint Y \psoddfigure 324 324 60 {Figure 6} {b_fit_p.ps} {
Fit parameters of P-wave wave functions
(a) $b$ parameter (b) $c$ parameter.
Points are labeled with crosses for ${}^3P_2$ mesons,
squares for ${}^1P_1$ mesons, diamonds for $N(5/2)$ baryons.
}\fi

\if\preprint Y \psoddfigure 324 324 60 {Figure 7} {b_fit_d.ps} {
Fit parameters of D-wave  wave functions
(a) $b$ parameter (b) $c$ parameter.
Points are labeled with crosses for ${}^3D_3$ mesons,
squares for ${}^1D_2$ mesons, diamonds for $N(7/2)$ baryons.
}\fi

\subhead{C. Moments of Wave Functions}
The $n$th moment of the meson wave function is
defined  in terms of $\psi(\vec r)$ as
$$\langle r^n\rangle = {\int r^2 dr\ ({r \over 2})^n f(r) ^2
\over {\int r^2 dr\ f(r)^2}}. \eqno(3.4)
$$
The factor of $1/2$ is included so that the second moment defined this way
(when appropriately weighted by quark charges)
reduces to the second moment of the quark charge distribution defined
through the form factor.

We determined the first and second moments of our meson wave functions
in two ways: first, we computed it directly from the data, by
performing a  single elimination
jacknife analysis, and second, we computed it using
the fitted form of wave functions. We consider the second method to be
 more reliable since in many cases the wave function is still large
at the edge of the simulation volume. Only a fit which includes image
effects can correctly reproduce the tail of the wave function.

In all cases and for both jacknife and
using the fitted radial wave function
 the values of two moments were independent of timeslice,
although the uncertainty increases with increasing $t$.
We display the first and second moments at $t=4$
in Figs. 8 and 9 respectively.
We see that while the two methods give rather similar results for
the $\kappa=.1300$ mesons, at smaller quark mass the discrepancy becomes
pronounced.
\if\preprint Y \psoddfigure  324 324 60 {Figure 8} {mom1.ps} {
First moment of P-wave and D-wave meson wave functions
(a) from a jacknife analysis and (b) from the fitted radial
wave functions. Points are labeled with crosses for ${}^3P_2$,
squares for ${}^1P_1$, diamonds for ${}^3D_3$, octagons for
${}^1D_2$ mesons, and fancy squares for the pseudoscalar (data from Ref.
\cite{TDMHSWAVE}).
}\fi

\if\preprint Y \psoddfigure  324 324 60 {Figure 9} {mom2.ps} {
Second moment of P-wave and D-wave meson wave functions
(a) from a jacknife analysis and (b) from the fitted radial
wave functions. Points are labeled as in Fig. 8.
}\fi

The P-wave wave functions are larger than the S-wave wave functions, and the
D-wave ones are larger still. Notice that  the diameter of
the wave function in the simulation is four times $\langle r \rangle$
of Eqn. (3.4),
so that the simulation volume we use would appear to be too small for D-wave
systems made of lighter quarks.

\head{IV. Spectroscopy}
We extracted masses from our data by fitting the correlation function
$C(t)$ of Eqn. (5) in the standard way, and looked at
``effective masses'' (local slope of $C(t)$) and fits to a range
$t_{min}$ to $t_{max}=n_t/2=8$.
All data are fit including the effects
of correlations at different times\rlap.\refto{DOITRIGHT}

As a general rule for selecting the best fit
value to present in a figure or table, we use  ``fit histograms.''  A
fit is represented by a rectangle centered on the best fit value for a
mass $\mu$,
with a width given by (twice) the uncertainty of the fit
 (i.e. $\mu \pm \Delta \mu$), and a height which is the
confidence level of the fit (to emphasize good fits) times the
number of degrees of freedom (to emphasize fits over big distance
ranges) divided by the statistical error on the parameter
 (to emphasize fits with
small errors).    The fit with the greatest
height is the one we quote.
This was the method  used to select the best mass in an earlier
S-wave spectroscopy calculation\rlap.\refto{HEMCGC}

\subhead{A. P-wave Spectroscopy}

Fig. 10 shows effective masses and fits to a range of $t$ values
for the meson and baryon data.

 \if\preprint Y \psoddfigure 522 486 -6  {Figure 10} {p_r_eff.ps}
{Spectroscopy of P-wave mesons and baryons. (a) meson effective masses,
(b) meson fits to a range, (c) baryon effective masses and (d)
baryon fits to a range.
The three bands (in order of increasing mass) correspond to $\kappa=0.152$,
0.145, and 0.130.
 In each meson group, the ${}^1P_1$ state
is labelled by a cross, the ${}^3P_0$ state by an octagon, the
${}^3P_1$ state by a square, and ${}^3P_2$ state by a diamond.
For baryons, the cross labels the $P_{1/2}$ state, the octagon
the $P_{3/2}$ state, and the square the $P_{5/2}$ state.}
\fi
All baryon masses at all $\kappa$ values appear to be consistent;
there is little sign of a drift of the masses with choice of fitting
range.  There  is no evidence of any fine structure splitting in any of
the $\kappa$ values. One cannot say whether this is due to a small
intrinsic splitting on the lattice, or whether all operators are
merely coupling strongly to the $j=5/2$ nucleon. In our extrapolations we
will make the latter assumption.

The $\kappa=.1300$ mesons also have stable, consistent fits. The
best fit values from histograms all begin at $t=2$.
  There is a hint of the appearance of fine structure
splitting in the multiplet, as shown in Fig. 11. The splitting
qualitatively resembles charmonium fine structure splitting,
with the ${}^3P_0$ state the lightest and the other states more
nearly degenerate. However, uncertainties are so large that
this probably should not be taken too seriously.
 These data seem to be limited by statistics.

 \if\preprint Y \psoddfigure 324 324 60  {Figure 11} {fine.ps}
{ Fine structure splitting in the P-wave meson multiplets
(a) $\kappa=0.130$,
(b) $\kappa=0.145$,
(c) $\kappa=0.152$.
}
\fi

The $\kappa=.1450$ data is noisier by about a factor of two.
 The ${}^3P_2$ state is heavier than the
${}^1P_1$.  The ${}^3P_0$ signal never stabilizes; while the fit
from $t_{min}=2$ is satisfactory from the point of view of chi-squared
(7.0 for 5 degrees of freedom) fits at increasing $t_{min}$'s produce
monotonically falling masses.
Fit histograms are shown in Fig. 12.
 The ${}^3P_1$ state is degenerate
with the ${}^3P_2$, but with large errors.
 \if\preprint Y \psoddfigure 522 486 -6  {Figure 12} {histp450.ps}
{``Fit histograms'' of $\kappa=.145$ P-wave mesons.
 (a) ${}^3P_2$ mesons,
 (b) ${}^3P_1$ mesons,
 (c) ${}^3P_0$ mesons,
 (d) ${}^1P_1$ mesons.
}
\fi

Finally, the $\kappa=.1520$ data share the same features as the
$\kappa=.1450$ data, with slightly larger uncertainties.
The P-wave lattice masses are listed in Table I.

\subhead{B. D-wave spectroscopy}
The D-wave data is noisier than the P-wave data (as expected).
Typical uncertainties for masses are about 0.08, four times the P-wave
value.
All masses appear to be asymptotic by $t_{min}=2-3$ and all signals
disappeared into the noise by $t=6$.
Because the data is so noisy, we could see no evidence for fine structure
 splitting in a multiplet.
Fig. 13 shows effective masses and fits to a range of $t$ values
for the meson and baryon data.
The masses are listed in Table II.
 \if\preprint Y \psoddfigure 522 486 -6  {Figure 13} {d_r_eff.ps}
{Spectroscopy of D-wave mesons and baryons. (a) meson effective masses,
(b) meson fits to a range, (c) baryon effective masses and (d)
baryon fits to a range.
The two bands (in order of increasing mass) correspond to $\kappa=0.145$
 and 0.130.
 In each meson group, the ${}^1D_2$ state
is labelled by a cross, the ${}^3D_1$ state by an octagon, the
${}^3D_2$ state by a square, and ${}^3D_3$ state by a diamond.
For baryons, the cross labels the $D_{1/2}$ state, the octagon
the $D_{5/2}$ state, and the square the $D_{7/2}$ state.}
\fi

\subhead{C. Comparison to Experiment}
It is difficult to convert these lattice numbers into reliable quantities which
can be compared to experiment. At $\beta=6.0$ one is far from the
scaling region. S-wave spectroscopy  with Wilson fermions
does not agree with experiment.
The quark hopping parameters we have used are very distant from the
zero quark mass value. We will glance at two comparisons, but
we have to say that with the quality of the signals they should
probably not be taken very seriously as other than qualitative observations.

  \if\preprint Y \psoddfigure 324 324 60  {Figure 14} {masses.ps}
{Masses of P-wave and D-wave hadrons as a function of hopping parameter.
Diamonds label ${}^3P_2$ mesons, squares ${}^3P_1$, octagons
${}^3P_0$, and crosses, ${}^1P_1$.
P-wave baryons are labeled with a fancy cross,
D-wave mesons are labeled with a fancy square and D-wave baryons with a burst.
}
\fi
The masses are shown in Fig. 14.
First, if we extrapolate all masses linearly  in $\kappa$
to $\kappa_c=0.1567$, we find $am({}^3P_2)=.93(3)$,  $am({}^1P_1)=.73(4)$,
 $am(N(5/2))=1.07(4)$, and the D-wave meson and baryon are at 1.18(10) and
1.93(20) respectively.  The proton mass at $\kappa_c$ from our S-wave
wave function study\refto{TDMHSWAVE} is $am(N)=0.55(1)$, so we have
$m({}^3P_2)/m(N) = 1.70(6)$, $m({}^1P_1)/m(N) = 1.33(8)$,
$m(N(5/2))/m(N) = 1.94(8)$, and the ratios of the D-wave meson and baryon
to the nucleon mass are 2.14(18) and 3.5(4).

Experimental data corresponding to these states are
$m(a_2)/m(N) =1.40$ or $m(f_2)/m(N) = 1.35$, $m(b1)/m(N)=1.31$,
$m(N(1675))/m(N)=1.78$, $m(\rho_3(1690))/m(N)=1.80$, and
$m(N(2220))/m(N)=2.36$.  The P-wave masses look qualitatively correct,
but the D-wave states (from which the extrapolation in $\kappa$ is
enormous) are too high.

As another comparison to experiment we can try to predict the
mass of the D-wave states in charmonium. To do this, we must
extrapolate in  $\kappa$ from our $\kappa=.1300$ data point to
the charm mass. We also need a value for the lattice spacing $a$, which
could vary by thirty per cent at this $\beta$ value depending on how it is
chosen.  We determine $\kappa$ and $a$ by taking  lattice determinations
of the ${}^3P_2$ and ${}^3S_1$ states at $\kappa=0.130$, 0.145, and
0.152 and extrapolating their masses linearly in $\kappa$.
(We use the data of Ref. \cite{DL} for the $\kappa=.1300$ vector meson).
We determine the lattice spacing by fitting the extrapolated masses
to the  $\psi(3095)$ and $\chi(3555)$ masses.  This gives
a charm hopping parameter of $\kappa=.1224$ and an inverse lattice spacing
of $1/a=1790$ MeV. (Note that determining the lattice spacing from
our proton mass data would give $1/a=1710$ MeV; from the rho meson,
$1/a=2264$ MeV.)
The extrapolated common D-meson mass is then
3.99(16) MeV, where the error is only from the extrapolation.
The ${}^3D_1$ $c \bar c$ state is at 3.77 GeV but its mass
is influenced by the nearby $D \bar D$ threshold.
 Model calculations\refto{DWAVE}
 of D-wave states (some of which are narrow since their
decays to $D \bar D$ are forbidden) give masses of 3.81 to 3.84 GeV.
At this value of the lattice spacing our ${}^3P_2-{}^3P_0$ mass splitting
at $\kappa=.1300$ is 63 MeV; in charmonium the corresponding
number is 145 MeV.

\head{V. Conclusions}
In retrospect, many
 aspects of the project could have been done better.
We should have completely reconstructed  the spin structure of
all the different states.
We should have recorded wave functions for all timeslices and not
just for big $t$'s. Then we could have used the wave functions
we determined in Sec. III as the input for $\psi_{test}(r)$ for
spectroscopy calculations.

It is clear that this program could be carried to arbitrarily high angular
momentum states. To do so will probably require very high statistics,
a more fine-grained lattice (since the lobe structure of the
angular part of the wave function becomes more pronounced) and a larger
simulation volume, since the size of the wave function grows with
angular momentum.

Note that the uncertainties on the  P-wave masses fell by about a factor of two
when the data set increased from 20 to 80 lattices. This suggests that
the calculations of P-wave meson masses are almost certainly limited only
by statistics. Meson
 P-wave spectroscopy still needs to improve its uncertainties
by another factor of two to four before it can begin to make a
serious comparison with experimental data, but we believe that this would
be an easy thing for any large scale spectroscopy simulation to do.
A reliable method for identifying specific baryon states remains to
be demonstrated.

\vfill\supereject
\subhead{Acknowledgements}
We would like to thank
C. Liu for providing us with a copy of his matrix inverter.
The computations described in this work were carried out at the
Pittsburgh Supercomputing Center. The work was supported by the U. S.
Department of Energy.

\endpage
\TABLEcap{I}{ P-wave meson and baryon masses in lattice units.}
\vskip-24pt
$$
{\vbox{\offinterlineskip\halign{
\vrule\fstrut\quad\hfil#\hfil\quad&&\fstrut\quad\hfil#\hfil\quad\cr
\dbline\notext

\vrule height 18pt depth 7pt width 0pt
$\kappa$ & state & mass \endrule
\sgline
0.1300  & ${}^3P2$ &    1.747(23)         \endrule
0.1300  & ${}^3P1$ &    1.759(22)         \endrule
0.1300  & ${}^3P0$ &    1.712(19)         \endrule
0.1300  & ${}^1P1$ &    1.760(21)         \endrule
0.1450  & ${}^3P2$ &    1.313(25)         \endrule
0.1450  & ${}^3P1$ &    1.322(74)         \endrule
0.1450  & ${}^3P0$ &    1.278(35)         \endrule
0.1450  & ${}^1P1$ &    1.210(37)         \endrule
0.1520  & ${}^3P2$ &    1.05(3)         \endrule
0.1520  & ${}^3P1$ &    0.96(3)         \endrule
0.1520  & ${}^3P0$ &    1.03(7)         \endrule
0.1520  & ${}^1P1$ &    0.88(4)         \endrule
\sgline
0.1300  & baryon &    2.56(2)         \endrule
0.1450  & baryon &    1.79(4)         \endrule
0.1520  & baryon &    1.28(4)         \endrule
\sgline
}}
}
$$
\TABLEcap{II}{Dwave meson and baryon masses in lattice units.}
\vskip-24pt
$$
{\vbox{\offinterlineskip\halign{
\vrule\fstrut\quad\hfil#\hfil\quad&&\fstrut\quad\hfil#\hfil\quad\cr
\dbline\notext

\vrule height 18pt depth 7pt width 0pt
$\kappa$ & state & mass \endrule
\sgline
0.1300  & meson &    2.00(6)         \endrule
0.1450  & meson &    1.54(5)         \endrule
0.1300  & baryon &    3.0(1)         \endrule
0.1450  & baryon &    2.4(1)         \endrule
\sgline
}}
}
$$
\if\preprint N
\figurecaptions
\item{1.}
Wave function of the
$\kappa=0.152$ $N(5/2)$ baryon in the plane $z=0$. The contours
show interpolated lines of constant real $\psi$ in
 multiples of 0.2 times the maximum value of $\psi$.
The shading shows the value of the wave function (black is most negative,
white most positive), interpolating from the original $16^2$ lattice
to a $64^2$ grid.

\item{2.}
Wave function of the $\kappa=.145$ ${}^3D_3$ meson in the plane $z=0$. The
 contours and shading are parameterized as in Fig. 1.

\item{3.}
Surfaces of constant absolute amplitude of the
 $\kappa=.145$  ${}^3D_3$ meson in
three dimensions, in fractions of its maximum: (a) $|\psi|= 0.2\psi_{max}$,
 (b) $|\psi|= 0.4\psi_{max}$ , (c) $|\psi|= 0.5\psi_{max}$,
 (d) $|\psi|= 0.6\psi_{max}$. The ``breaks'' in the surfaces in (a)
and (b) occur when the surfaces intersect the edge of the
simulation volume.

\item{4.}
Radial wave function of the $\kappa=.145$  ${}^3P_2$ meson.
Crosses show all points; squares are points along the  $x$ axis
used to fit the wave function (fit is the line).

\item{5.}
Radial wave function of the $\kappa=.145$  ${}^3D_3$ meson.
Crosses show all points; squares are points
used to fit the wave function (fit is the line).

\item{6.}
Fit parameters of P-wave wave functions
(a) $b$ parameter (b) $c$ parameter.
Points are labeled with crosses for ${}^3P_2$ mesons,
squares for ${}^1P_1$ mesons, diamonds for $N(5/2)$ baryons.

\item{7.}
Fit parameters of D-wave  wave functions
(a) $b$ parameter (b) $c$ parameter.
Points are labeled with crosses for ${}^3D_3$ mesons,
squares for ${}^1D_2$ mesons, diamonds for $N(7/2)$ baryons.

\item{8.}
First moment of P-wave and D-wave meson wave functions
(a) from a jacknife analysis and (b) from the fitted radial
wave functions. Points are labeled with crosses for ${}^3P_2$,
squares for ${}^1P_1$, diamonds for ${}^3D_3$, octagons for
${}^1D_2$ mesons, and fancy squares for the pseudoscalar (data from Ref.
\cite{TDMHSWAVE}).

\item{9.}
Second moment of P-wave and D-wave meson wave functions
(a) from a jacknife analysis and (b) from the fitted radial
wave functions. Points are labeled as in Fig. 8.

\item{10.}
Spectroscopy of P-wave mesons and baryons. (a) meson effective masses,
(b) meson fits to a range, (c) baryon effective masses and (d)
baryon fits to a range.
The three bands (in order of increasing mass) correspond to $\kappa=0.152$,
0.145, and 0.130.
 In each meson group, the ${}^1P_1$ state
is labelled by a cross, the ${}^3P_0$ state by an octagon, the
${}^3P_1$ state by a square, and ${}^3P_2$ state by a diamond.
For baryons, the cross labels the $P_{1/2}$ state, the octagon
the $P_{3/2}$ state, and the square the $P_{5/2}$ state.

\item{11.}
Fine structure splitting in the P-wave meson multiplets
(a) $\kappa=0.130$,
(b) $\kappa=0.145$,
(c) $\kappa=0.152$.

\item{12.}
``Fit histograms'' of $\kappa=.145$ P-wave mesons.
 (a) ${}^3P_2$ mesons,
 (b) ${}^3P_1$ mesons,
 (c) ${}^3P_0$ mesons,
 (d) ${}^1P_1$ mesons.

\item{13.}
Spectroscopy of D-wave mesons and baryons. (a) meson effective masses,
(b) meson fits to a range, (c) baryon effective masses and (d)
baryon fits to a range.
The two bands (in order of increasing mass) correspond to $\kappa=0.145$
 and 0.130.
 In each meson group, the ${}^1D_2$ state
is labelled by a cross, the ${}^3D_1$ state by an octagon, the
${}^3D_2$ state by a square, and ${}^3D_3$ state by a diamond.
For baryons, the cross labels the $D_{1/2}$ state, the octagon
the $D_{5/2}$ state, and the square the $D_{7/2}$ state.

\item{14.}
Masses of P-wave and D-wave hadrons as a function of hopping parameter.
Diamonds label ${}^3P_2$ mesons, squares ${}^3P_1$, octagons
${}^3P_0$, and crosses, ${}^1P_1$.
P-wave baryons are labeled with a fancy cross,
D-wave mesons are labeled with a fancy square and D-wave baryons with a burst.

\endfigurecaptions
\fi
\references

\refis{TDMHPWAVE}
T. DeGrand and M. Hecht, \journal Phys. Lett., B275, 435, 1992.

\refis{TDMHSWAVE}
M. W. Hecht and T. DeGrand, Colorado preprint COLO-HEP-277, 1992.

\refis{FNAL}
A.  El-Khadra, G. Hockney, A. Kronfeld and P. Mackenzie, Fermilab preprint
FERMILAB-PUB-91/354-T, 1991;
P. Mackenzie, Fermilab preprint FERMILAB-CONF-92/09-T, 1991;
A. El-Khadra, Fermilab preprint FERMILAB-CONF-92/10-T, 1991.

\refis{APE}
P. Bacilieri, et. al., \journal Phys. Lett., 214B, 115, 1988,
\journal Nucl. Phys., B317, 509, 1989.

\refis{APETWO}
S. Cabasino, et. al.,  \journal Phys. Lett., 258B, 195, 1991.


\refis{LIU}
C. Liu, in
 the Proceedings of Lattice '90,
{\sl Nucl. Phys.} {\bf B (Proc. Suppl) 20},  (1991) 149.

\refis{OVERRELAX}
J. E.~Mandula and M. C.~Ogilvie, \pl B248, 156, 1990.

\refis{LEPAGET}
B. Thacker and G. P.  Lepage, \journal Phys. Rev., D43, 196, 1991.

\refis{KOKKEDDEE}
J. Kokkedee, ``The Quark Model,'' (W. A. Benjamin,  New York, 1969).

\refis{ALLWF}
Wave functions computed in a smooth gauge were first introduced by
Velikson and Weingarten, \journal Nucl. Phys., B249, 433, 1985
and by
 S. Gottlieb, in ``Advances in Lattice Gauge Theory,''
D. Duke and J. Owens, eds. (World Scientific, 1985).
Other recent uses, for heavy-light quark systems are described by E. Eichten,
in
 the Proceedings of Lattice '90, U. Heller, A. Kennedy and S. Sanielevici, eds.
{\sl Nucl. Phys.} {\bf B (Proc. Suppl) 20},  (1991) 475
and
C. Bernard, J. Labrenz, and A. Soni, ibid., 488.
A gauge-invariant formalism has been described by
 M.-C.~Chu, M.~Lassia and J. W.~Negele,  \journal Nucl. Phys., B360, 31,  1991.

\refis{RELAX}
F. Brown and T. Woch, \journal Phys. Rev. Lett., 58, 2394, 1987;
M. Creutz, \journal Phys. Rev., D36, 55, 1987.
For a review, see S. Adler in
the Proceedings of the 1988 Symposium on
Lattice Field Theory, A. Kronfeld and P. Mackenzie, eds.,
{\sl Nucl. Phys.} {\bf B (Proc. Suppl) 9},  (1989) 437.

\refis{KPQ}
A. Kennedy and B. Pendleton, \journal Phys. Lett., 156B, 393, 1985.

\refis{DL}
T. DeGrand and R. Loft, \journal  Phys. Rev.,  D39, 2678, 1989.

\refis{DOITRIGHT}
For a good introduction to error analysis see
D. Toussaint, in
``From Actions to Answers--Proceedings of the 1989 Theoretical Advanced
Summer Institute in Particle Physics,'' T. DeGrand and D. Toussaint, eds.,
(World, 1990).

\refis{LEPAGE}
See
G. P.~Lepage, in
``From Actions to Answers--Proceedings of the 1989 Theoretical Advanced
Summer Institute in Particle Physics,'' T. DeGrand and D. Toussaint, eds.,
(World, 1990).

\refis{HEMCGC}
K.~Bitar et al., \prl 65, 2106, 1990, \prd 42, 3794, 1990.

\refis{DWAVE}
Compare Eichten, et. al., \journal Phys. Rev., D21, 203, 1980 (3.81 GeV);
W. Buchm\"uller and S-H. Tye, \journal Phys. Rev., D24, 132, 1981 (3.81 GeV);
W. Kwong, J. Rosner, C. Quigg, \journal Ann. Rev. Nucl. Part. Sci., 37, 325,
1987 quote $m({}^3D_2)=3.81$ GeV,  $m({}^3D_3)=3.84$ GeV,
 $m({}^1D_2)=3.82$ GeV from  the model of
P. Moxhay and J. Rosner, \journal Phys. Rev., D28, 1132, 1983.

\endreferences
\endit
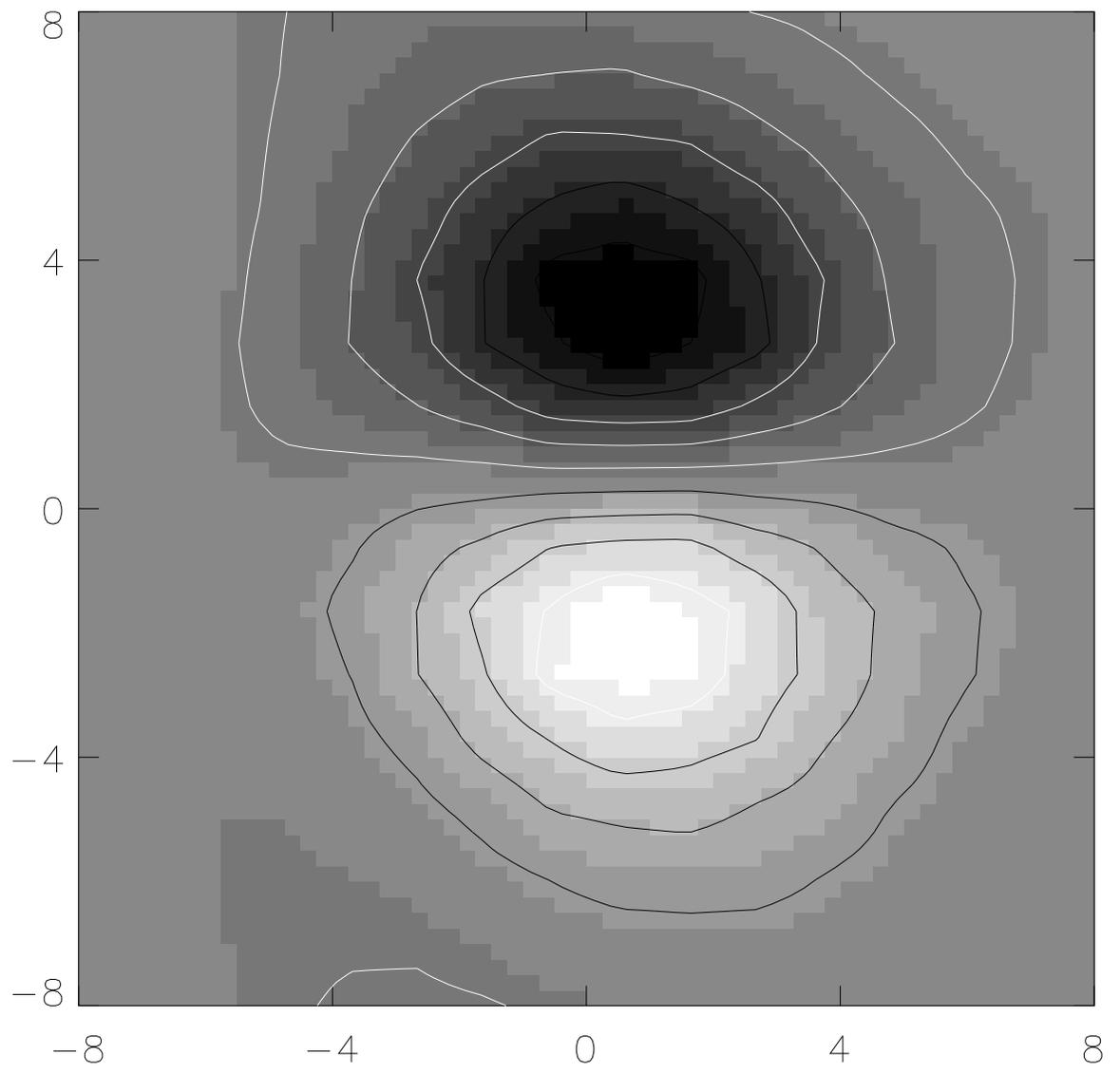
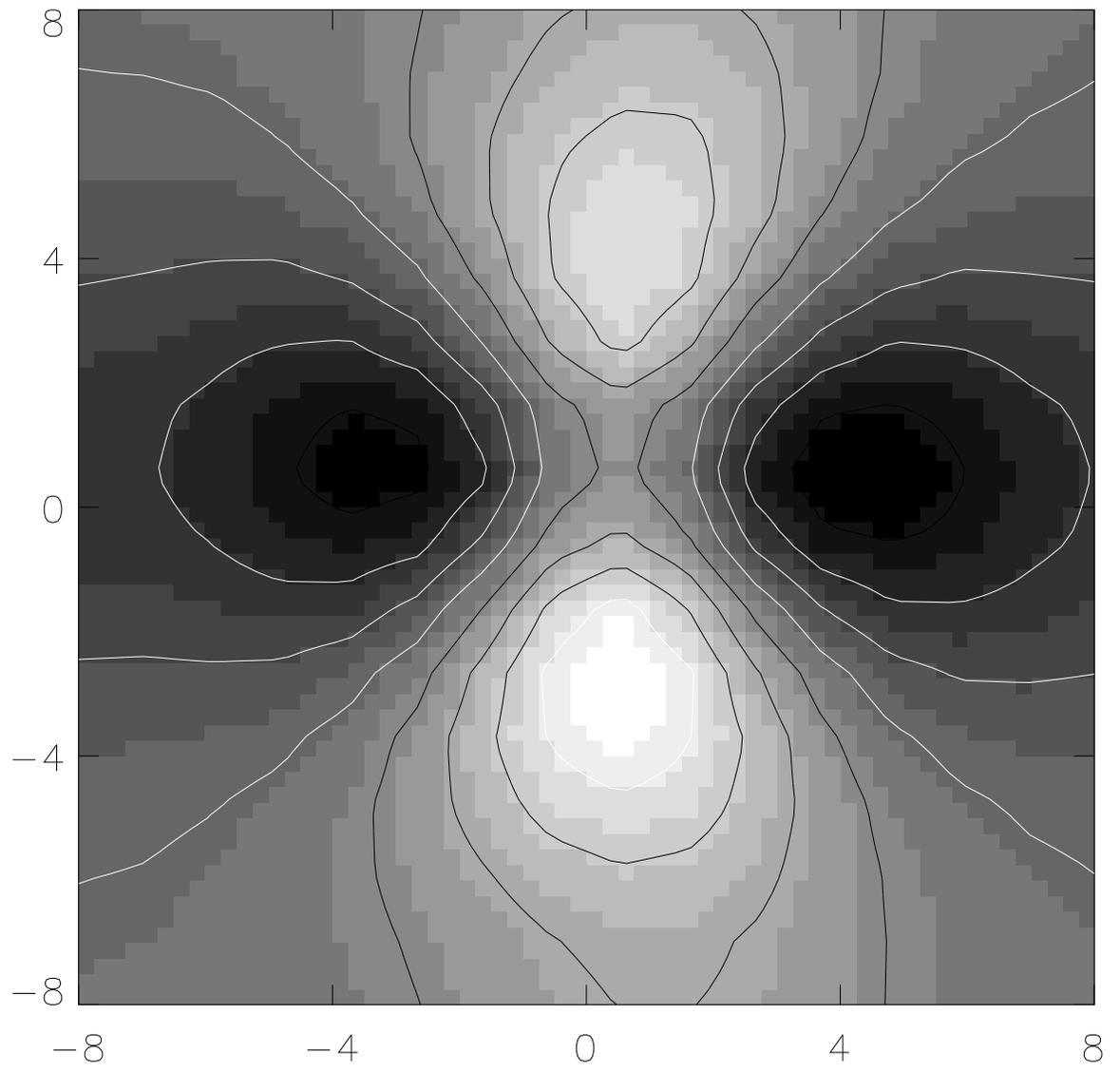
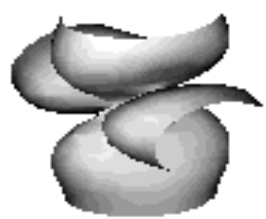
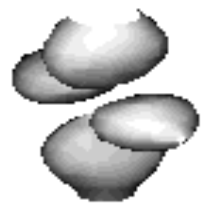
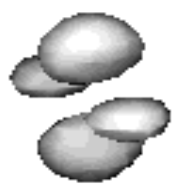
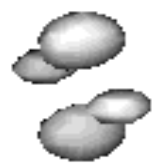
